\documentclass[twocolumn]{aastex631}
\usepackage{amsmath}

\usepackage{enumitem}

\begin{document}

\title{A Linearly Polarized Merger Shock Down to 550 MHz: A uGMRT Study of the Merging Cluster Abell 746}

\author[0000-0002-0786-7307]{Arpan Pal}
\affiliation{National Centre for Radio Astrophysics, Tata Institute of Fundamental Research, S. P. Pune University Campus, Ganeshkhind, Pune, 411007}

\author[0000-0003-1449-3718]{Ruta Kale}
\affiliation{National Centre for Radio Astrophysics, Tata Institute of Fundamental Research, S. P. Pune University Campus, Ganeshkhind, Pune, 411007}

\author[0000-0002-0587-1660]{Reinout J. van Weeren}
\affiliation{Leiden Observatory, Leiden University, PO Box 9513, NL-2300 RA Leiden, The Netherlands}

\begin{abstract}
Radio relics, arc-like polarized sources with highly aligned magnetic fields, are typically found on the outskirts of merging galaxy clusters. The magneto-ionic media responsible for the significant coherence observed in radio relics remain poorly understood. Low-frequency measurements of radio relics are essential for constraining depolarization models, which provide crucial insights into the magnetic field distribution. However, these measurements are challenging due to the emission properties and interferometer systematics. We have detected polarization signals from the northwest radio relic in Abell 746 at 650 MHz with the upgraded Giant Meterwave Radio Telescope, marking the first-ever detection of polarization from radio relics below 1 GHz. At this frequency, the average Rotation Measure (RM) corrected magnetic fields align well with shock radio emission, typical of radio relics. The fractional polarization at 650 MHz is $\sim 18\pm4$\ \%. Our results indicate that a single internal depolarization model cannot explain the observed depolarization spectra, suggesting a non-uniform magnetic field distribution or complex contribution of different polarized regions in the radio relic. Our detection of polarization signals at 650 MHz reveals critical insights into radio relic magnetic field structures, offering a low-frequency approach to understanding ICM magnetic fields in merging galaxy clusters.

\end{abstract}

\keywords{Diffuse radio emission, polarization of radio relics, galaxy cluster merger, low-frequency polarization}

\section{Introduction} \label{sec:intro}
Galaxy cluster collisions generate low Mach number (M $\sim 2-5$; e.g., \citealp{2012ApJ...748....7M,2022MNRAS.514.1477R,2025ApJ...978..122L})  shocks and turbulence in the intra-cluster medium (ICM), a hot ($\sim 10^8$ K)
diffuse plasma that permeates 
the cluster volume. The relatively mild shocks and turbulence act as mechanisms to convert immense gravitational potential energy into thermal energy.

During these merger events, Comsic Rays (CR) \citep{2011A&A...527A..99E} undergo (re)acceleration at shock fronts (e.g., \citealp{1977DoSSR.234.1306K};\citealp{1977ICRC...11..132A, 1978MNRAS.182..147B, 1987PhR...154....1B}), and
the ICM contains large-scale magnetic fields \citep{2010A&A...513A..30B, 2022A&A...665A..71O}, whose properties remain challenging to measure precisely due to the low strengths (0.1 - 10 $\mu$G) and complex topology across megaparsec scales. Faraday rotation measures only provide line-of-sight components weighted by electron density, while synchrotron emission analysis suffers from degeneracy between field strength and relativistic electron density. Additionally, projection effects introduce significant uncertainties in interpreting observational data. These magnetic fields play a crucial role in governing plasma dynamics through magnetohydrodynamic (MHD) processes such as turbulent cascades, bulk flows, and shear motions within the ICM, while also significantly influencing the formation of large-scale cosmic structures \citep{2001PhR...348..163G, 2012SSRv..166....1R}. Through their interaction with the magnetic fields, the CR electrons (CRes) produce synchrotron radiation that is observable at radio wavelengths. These radio signatures, known as radio relics, trace the merger shocks and provide unique laboratories for investigating the poorly understood magnetic fields within galaxy clusters and their complex interactions with shock.

Radio relics display remarkable polarization properties at frequencies above 1 GHz, with average fractional polarization reaching $\geq 20-30\%$ (e.g., \citealp{2010Sci...330..347V, 2012MNRAS.426.1204K,2017A&A...600A..18K, 2022A&A...666A...8S, 2025ApJ...979....4P}). These structures exhibit coherent electric field position angles (EVPAs) aligned with shock normals across megaparsec scales (e.g., \citealp{2010Sci...330..347V, 2021ApJ...911....3D,2025ApJ...979....4P}), suggesting highly organized magnetic fields in their emission regions.
The origin of this magnetic field alignment presents a fundamental puzzle: whether it stems from pre-existing large-scale fields or results from shock compression of turbulent fields \citep{2019MNRAS.490.3987W}. Magneto-hydrodynamical simulations \citep{2019MNRAS.490.3987W, paola21} suggest that the polarization fraction is high at the shock front and decreases in the downstream region if the ICM is turbulent. In contrast, it increases in the downstream regions if the medium is uniform. However, in simulated radio relics \citep{2019MNRAS.490.3987W}, though local alignments were observed, above scales of 500 kpc, the global alignment of polarization vectors was not seen.

Simulations (\citealp{2018MNRAS.474.1672V, 2019MNRAS.490.3987W, paola21}) predict non-Gaussian rotation measure (RM) distributions in radio relics. While some observations \citep{https://www.aanda.org/articles/aa/pdf/2022/10/aa44179-22.pdf} validate the non-Gaussianity, high-resolution, wideband polarization studies show significant spatial variations in RM across relics \citep{2021ApJ...911....3D,2022A&A...657A...2R}, which adds complexity when modelling the overall RM distribution of radio relics. It is worth mentioning that for the Sausage radio relic, the single-component depolarization is consistent with a Gaussian distribution in the RM \citep{2021ApJ...911....3D}. The RM, which quantifies the rotation of polarization angle due to propagation through a magneto-ionic plasma, provides crucial insights into the structure of cluster magnetic fields. These non-Gaussian distributions suggest complex magnetic field configurations rather than uniform fields.

Radio relics exhibit frequency-dependent polarization properties, with their linear fractional polarization decreasing at longer wavelengths—a phenomenon known as depolarization. The amount of this depolarization strongly correlates with the magnitude and direction of magnetic fields in the emission region. Given the measured and simulated standard deviation of the RM distributions ($\sigma_{\textrm{RM}} = 10-100 \textrm{ rad m}^{-2}$; e.g., \citealp{2019MNRAS.490.3987W, https://arxiv.org/abs/2108.06343, https://arxiv.org/abs/2207.00503, https://arxiv.org/abs/2102.06631}) in the radio relics, polarization observations below 1 GHz allow us to weigh different depolarization models as they significantly deviate from each other in the low-frequency regime. But from the extrapolation of the higher frequency linear fractional measurements, the radio relics were thought to be severely depolarized below 1 GHz.

\begin{figure*}[ht!]
\includegraphics[width=\textwidth]{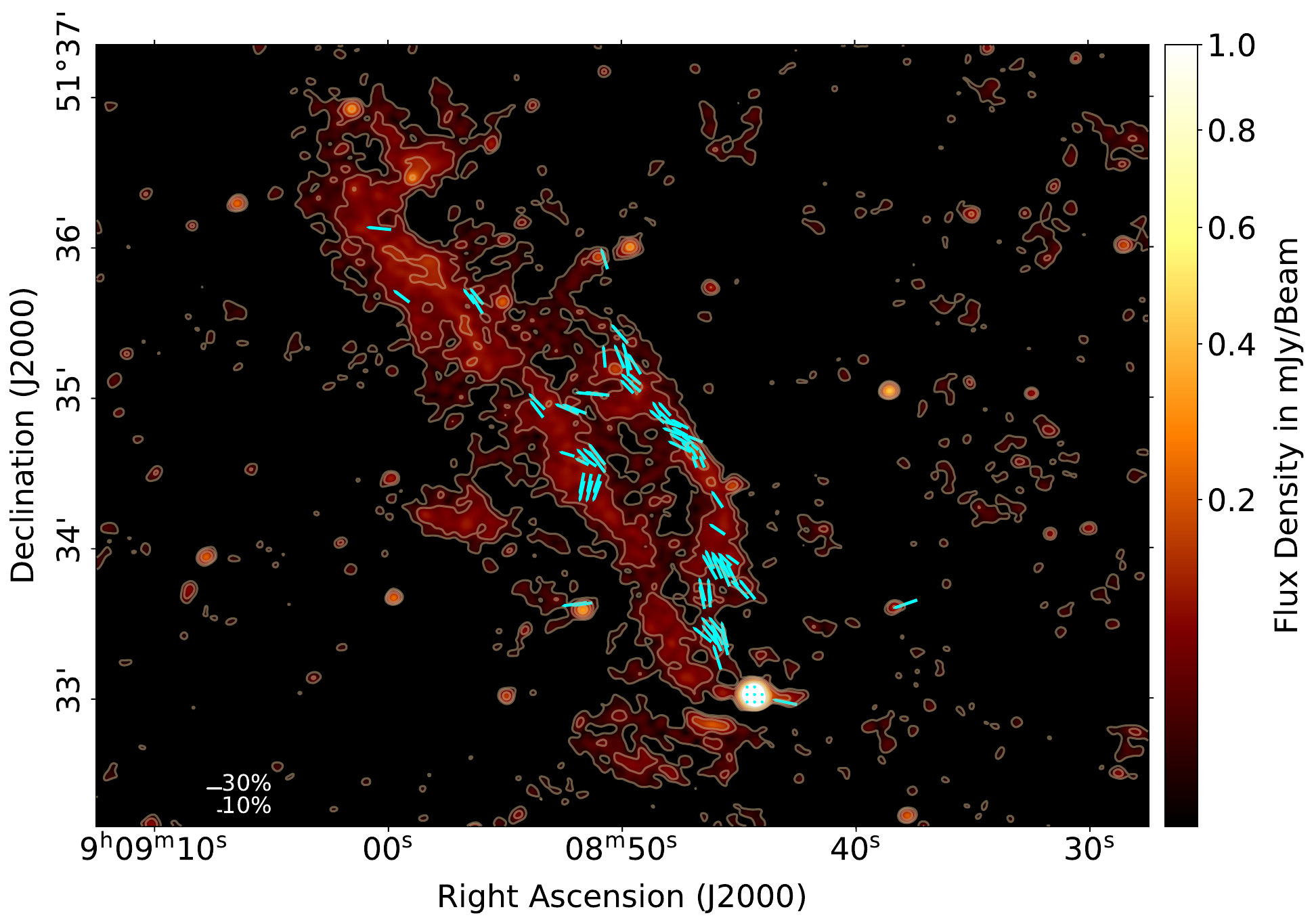}
\caption{uGMRT 650 MHz \( 4.6\arcsec \times 4.2\arcsec \) intensity in color scale. The white contours start at \( 5\sigma \), with successive levels increasing by a factor of \( \sqrt{2} \). The cyan vectors represent the magnetic fields (polarization angles, rotated by $90^\circ$), corrected for the average RM and the vector lengths are proportional to the linear fractional polarization. To show the distribution of the magnetic fields, 4 pixels are averaged while overplotting the vectors.}
\label{G166_pol_20}
\end{figure*}

Efforts to measure linear polarization at low frequencies (below 1 GHz) have also resulted in non-detections \citep{2011A&A...526A...9B,2011A&A...525A.104P, https://doi.org/10.1093/pasj/psv082}. These challenges arise primarily from the limited resolution and sensitivity of earlier interferometers, compounded by difficulties in sample selection (e.g., \citet{2011A&A...526A...9B,2011A&A...525A.104P}) and observing at very low-frequencies like 350 MHz. The ICM significantly contributes to the depolarization of linear polarization signals, particularly in systems with complex source morphologies. Such morphologies are often associated with intricate ICM structures, which increase susceptibility to depolarization \citep{https://doi.org/10.1093/pasj/psv082}. Modern radio interferometers, with their enhanced sensitivity and resolution, enable new investigations of sub-GHz polarization in radio relics. Double radio relic systems, where both shock fronts are discernible, present the most promising targets for detecting polarization at lower frequencies. These systems are believed to result from mergers occurring predominantly in the plane of the sky, benefiting from a favourable geometry that minimizes the line-of-sight projection through the ICM, thereby minimizing polarization loss due to orientation effects \citep{1998A&A...332..395E}.

Abell 746 is a moderately high-mass ($5.34\pm 0.4 \times 10^{14}\text{M}_\odot$; \citealp{0.1051/0004-6361/201525830}) merging cluster at redshift 0.214 \citep{2024ApJ...962..100H}. The cluster hosts four radio relics—Northwest (NW) \citep{https://arxiv.org/pdf/1107.5597, 2022A&A...660A..78B, https://arxiv.org/abs/2309.01716}, R1, R2, and R3—and an elongated radio halo \citep{2022A&A...660A..78B,https://arxiv.org/abs/2309.01716}. Previously, only the NW relic showed polarization in 1.5 GHz VLA observations \citep{https://arxiv.org/abs/2309.01716}. We observed the system with uGMRT band 4 (550-750 MHz, P.I. Arpan Pal, Proposal Code: 45\_087) to explore the low-frequency linear polarization properties of the system. Our uGMRT observations reveal linear polarization in the NW radio relic at 650 MHz. This paper presents our observational details and analysis procedure in Section \ref{sec:dataanalysis}, followed by the results and discussion on the magnetism of radio relics in Section \ref{sec:results}.

\section{Observation details and Data Analysis}\label{sec:dataanalysis}
We observed Abell 746 using uGMRT band 4 (550 - 750 MHz) on 28th November 2023 (proposal code 45\_087). We used 4096 channels over 200 MHz bandwidth with a total on-source time of 5.5 hours. 3C286 and 3C147 were observed as primary calibrators for gain, delay, and bandpass. The secondary gain calibrator was 0834+555. Additionally, 3C286 was used to calibrate the R-L phase and estimate cross-hand delays, while 3C147 was used to determine polarization leakage between the two polarizations. The uGMRT, as well as the legacy GMRT, which received the signal through a circularly polarized feed at band 4, captured a flipped response due to the reflection by the parabolic reflector \citep{2020arXiv200408542D, https://arxiv.org/abs/2310.04335}. Conventional polarization calibration in Common Astronomy Software Applications (CASA;  \citealp{2022PASP..134k4501C}) only rotated complex plane signals without addressing reflection while reaching the feed. Polarization calibration with flipped feeds incorrectly fixed polarizer axes relative to the sky.
To rectify this, we implemented a header swap: manually reconfiguring Stokes keywords from RR, RL, LR, LL to LL, LR, RL, RR.\par

We used radio interferometric data analysis software CASA 6.5 for data analysis. First, the bad antennas and time ranges were flagged. Then a round of automated flagging using CASA task `flagdata' with mode `tfcrop' was executed on all the fields with conservative thresholds of $6\sigma$ across time and frequency plane to flag obvious outliers. Using the whole scan of the 3C147 and 3C286, the initial gain phases were estimated. Using the initial gain phases, parallel-hand residual delays were estimated using the whole frequency range with the CASA task `gaincal' with `gaintype' `K'. With the estimated delays, the bandpasses were computed using the scans of both 3C286 and 3C147 using CASA task `bandpass'. With the bandpass, for both the 3C147 and 3C286, refined gain amplitude and phases were computed using `gaincal' with `gaintype' `G' and `calmode' `ap'. The secondary phase calibrator's gains were also estimated assuming a 1 Jy response. The amplitudes were then scaled by comparing them to the absolute flux calibrator (3C147, 3C286) solutions using the CASA task `fluxscale'. \par
With the gains properly computed, we then moved onto estimating the cross-hand corruptions. We found out from our experience that an extrapolation of the higher frequency (up to 1 GHz) linear fractional polarization and polarization angle measurements to lower frequencies did not necessarily capture the actual response of the polarized calibrator 3C286. Hence, the response of the polarized calibrators had to be measured for the relevant frequency band. As a separate exercise, we observed 3C286, our main polarization angle calibrator in regular cadence and with different parallactic angles. An updated full polar model of the 3C286 was then estimated using different combinations of unpolarized and polarized calibrators. The model was then cross-matched with MeerKAT UHF band measurements (personal communications with Preshanth Jagannathan and Ben Hugo \citep{hugo2024absolute}). The details of the analysis and the model will be discussed in a separate paper (Pal et al. in Prep.).\par
The full polarization model of 3C286 was then supplied to the CASA using `setjy'. First, the cross-hand delays were estimated using the CASA task `gaincal' with mode `kcross' using the 3C286 scan. Then leakages were estimated using the 3C147 scans assuming an unpolarized response in the 550-750 MHz range. With all the information, the cross-hand phase corruptions were computed on 3C286. The CASA tasks used for leakage and cross-hand phase corrections were `polcal' with mode `Df' and `Xf' respectively.\par
All the solution tables were applied to both calibrators and target source to achieve a preliminary full-polarization calibration. CASA task, flagdata with modes `tfcrop' and `rflag' were then used to flag the unwanted radio frequency interference (RFI). After the flagging, we re-did a full-polarization calibration of the visibilities following the exact same methodology. Now, as most of the RFIs were removed, the re-calibration allowed us to refine the solutions of all the relevant quantities.
Antennas that had a leakage of more than $20\%$ were flagged at the first stage. Antenna `W03' was the worst of all antennas with a leakage of ~35-40 \%. The other antennas had an average leakage of 5-15 \%. 
After the second round of calibration, the data were again flagged using CASA task `flagdata' mode `rflag' with baseline-based ranges, where the uv-plane was divided into independent segments of 500$\lambda$ step size to avoid unnecessarily flagging the short baselines. Afterwards, the desired source was split into a separate measurement file with frequency averaging by a factor of 10 channels, reducing the data from 4096 to approximately 410 channels. We ensured the averaging did not introduce the bandwidth-smearing and intra-channel depolarization effects as much as possible. The source was then imaged using the CASA task `tclean' with a robust value of 0 and nterms of 2. After several rounds of phase-only self-calibration of the parallel hand visibilities, a couple of amplitude-phase self-calibrations were also performed on the parallel hand visibilities. We did not self-calibrate leakage and cross-hand phase offset in the self-calibration loops. At the final stage, all the Stokes parameters were imaged using `tclean' with the robust value of 0 and nterms of 2 over the full band of 200 MHz. Then every image was primary beam corrected using \href{https://github.com/arpan-52/Finalflash}{Finalflash}. We also looked at the off-axis leakage terms to make sure of the off-axis contributions of the primary beams (Pal et al. in Prep.). The relic had a maximum distance of $3\arcmin$ from the pointing centre, implying non-significant off-axis leakages. Using the primary beam corrected IQU images, we selected only those regions which are detected over \(3\sigma\) in total intensity and \(5\sigma\) in linear polarization intensity to construct the final linear fractional polarization and polarization angle images.

\begin{figure*}[ht!]
\includegraphics[width=\textwidth]{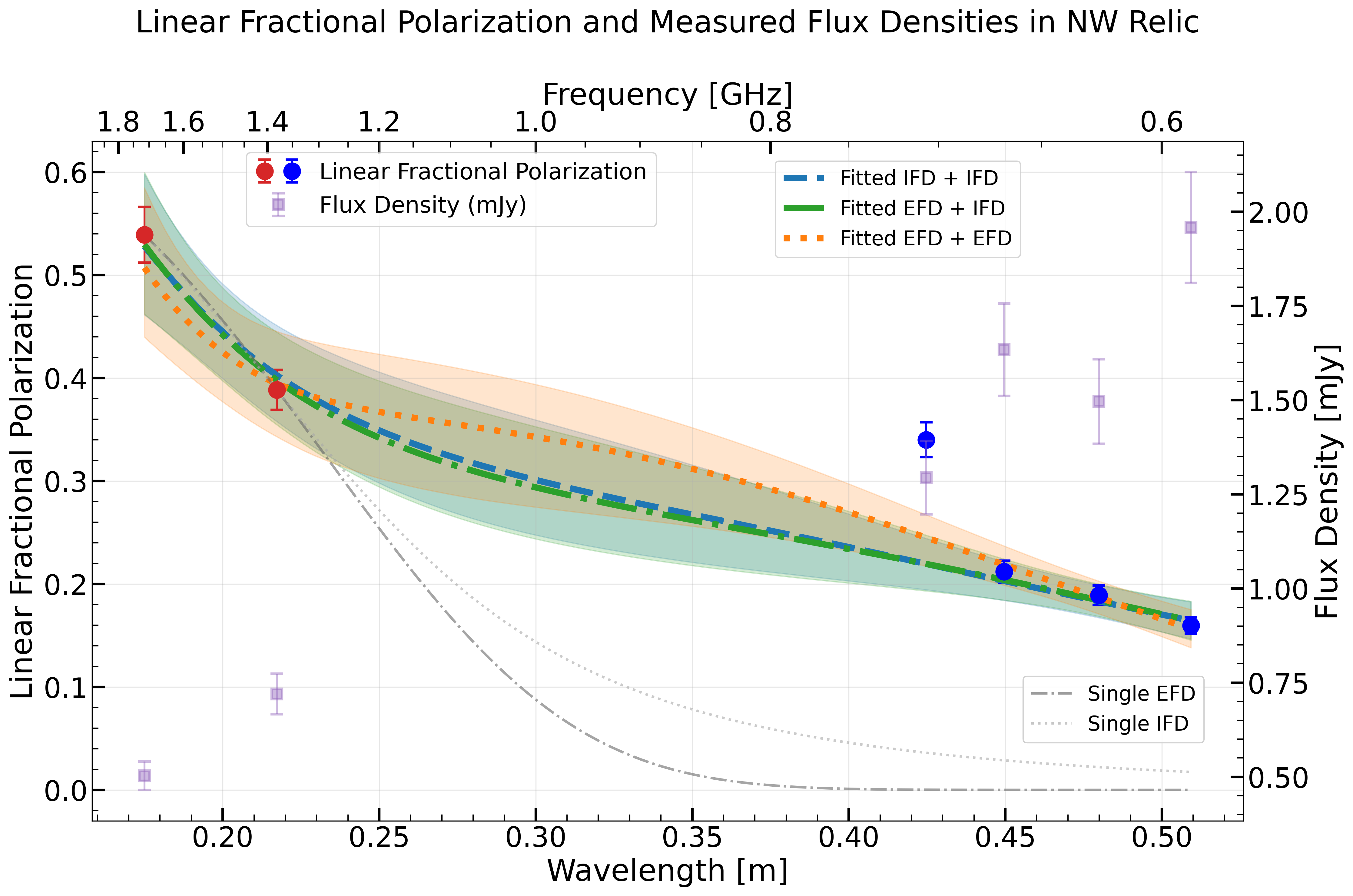}
\caption{The left Y-axis represents the linear fractional polarization, while the right Y-axis represents the flux densities of the region selected 
from the radio relic using the total intensity and polarized intensity cutoffs, as mentioned in \ref{sec:dataanalysis} used for calculating the linear fractional polarization. The red circles represent the linear fractional polarization measurements with the Westerbork Synthesis Radio Telescope (WSRT) at 1.71 and 1.38 GHz. The blue circles represent the linear fractional polarization measurements with the uGMRT band 4 (550-750 MHz). The orange dotted, blue dashed, and green dash-dotted lines represent the fitted External Faraday Dispersion-Internal Faraday Dispersion (EFD-IFD), Internal Faraday Dispersion-Internal Faraday Dispersion (IFD-IFD), and External Faraday Dispersion-External Faraday Dispersion (EFD-EFD) models, respectively, fitted to the linear polarization data. The filled regions represent the 1$\sigma$ uncertainties over the fitted models. The purple boxes represent the flux densities. The grey dotted and dash-dot lines represent single component EFD and IFD models derived from the 1.38 and 1.71 MHz WSRT measurements.}
\label{G166_mulpol}
\end{figure*}

The linear fractional polarization is defined as:

\begin{equation}
\frac{\sqrt{Q^2 + U^2}}{I},
\end{equation}

with its uncertainty given by:

\begin{equation}
\sigma_p = \frac{1}{I \sqrt{Q^2 + U^2}} \sqrt{Q^2 \sigma_Q^2 + U^2 \sigma_U^2} + \frac{\sqrt{Q^2 + U^2}}{I^2} \sigma_I,
\end{equation}

where \( \sigma_I \), \( \sigma_Q \), and \( \sigma_U \) represent the uncertainties in the flux densities of \( I \), \( Q \), and \( U \), respectively. These uncertainties are calculated as:

\begin{equation}
\Delta F = \sqrt{(0.1 F)^2 + (N_{\text{beams}} \times \text{r.m.s.})^2},
\end{equation}

where \( F \) is the flux density in the region of interest, \( N_{\text{beams}} \) is the number of beams covering the region, and \(\text{r.m.s.}\) is the sensitivity of the relevant image. While calculating the polarized intensity $\text{I}_\text{p} = \sqrt{\text{Q}^2 + \text{U}^2}$, we have corrected for Ricean Bias using the formula 
\begin{equation}
    \text{I}_{p}^{\text{corr}} = \text{I}_{p}^{\text{obs}}\sqrt{ (1 - \frac{\sigma_{\text{I}_p}}{\text{I}_{p}^{\text{obs}}})^2}
\end{equation}
where the $\text{I}_{p}^{\text{corr}}$ is the corrected polarized intensity, $\text{I}_{p}^{\text{obs}}$ is the observed polarized intensity, and $\sigma_{\text{I}_p}$ is the standard deviation in the polarized intensity image following \citet{1974ApJ...194..249W}.
\par
We then created a Stokes QU spectral cube with 20 subbands each subband spanning 9.76 MHz of the data. Due to the faint nature of the emission and expected low linear fractional polarization, the radio relic was not detected in polarization in any of the subbands. Even with 8 subbands across the 200 MHz bandwidth, the relic remained undetected.

To further improve the signal-to-noise ratio, we reduced the spectral resolution by combining the data into four subbands, thereby increasing the bandwidth and sensitivity within each subband. While the broader bandwidth reduced the ability to resolve RM signatures, the radio relic got detected in the subbanded QU images. Since RM-synthesis was not possible due to insufficient frequency sampling , we applied the average RM value of $-10 \text{rad m}^{-2}$ from \citep{https://arxiv.org/abs/2309.01716} to de-rotate the polarization vectors.

These polarimetric measurements were supplemented with 1.38 and 1.71 GHz measurements from WSRT. The 1.38 GHz image was already published in \citet{https://arxiv.org/pdf/1107.5597}. Though the 1.71 GHz measurements were not published in the paper but the data reduction followed the same steps. The polarization data reduction follows the procedure outlined in \citet{2012A&A...546A.124V}.

To compare the fractional polarization across frequencies, we first smoothed all the WSRT and uGMRT images to a common resolution of $25\arcsec$ as this is the highest resolution the WSRT images allowed. A \(3\sigma\) cutoff was applied to the Stokes \(I\) images and a \(5\sigma\) cutoff to the linearly polarized intensity in all datasets. We then identified the regions that were polarized across all frequencies by comparing the WSRT and uGMRT images. Finally, we examined the variation in linear fractional polarization within this common region to constrain a depolarization model for the NW radio relic.

As in lower frequencies, the ionospheric RM contributions affect our ability to calibrate the polarization angles. We have not corrected for the ionospheric RM in our final images. Using the total electron column-density measurements from the International GNSS Service (\href{https://cddis.nasa.gov/}{IGS}) and magnetic fields from International Geomagnetic Reference Field (IGRF \footnote{\url{https://github.com/zzyztyy/pyIGRF}}; \citealp{2021EP&S...73...49A}) measurements, we measure an average Rotation Measure (RM) estimate of 3.5 rad.m$^{-2}$ throughout the span of the observation.

\section{Results and Discussion}\label{sec:results} In the deep 650 MHz images, the NW relic exhibits an average linear fractional polarization of approximately $18 \pm 4\%$ within the regions selected based on total intensity and polarized intensity thresholds. The polarization fraction varies across the shock surface, reaching $\sim 35\%$ at its outer edge. Among the whole radio relic, the Western edge shows the strongest polarization while the downstream region becomes increasingly depolarized (Fig. \ref{G166_pol_20}), similar to measurements at higher frequencies (Fig. \ref{compare}, and \citealp{https://arxiv.org/abs/2309.01716}) This is possibly due to growing turbulence away from the shock front that randomizes the initially ordered magnetic field \citep{2018ApJ...865...24D}. The anisotropy of the downstream magnetic field governs the level of depolarization in the post-shock region \citep{2018ApJ...865...24D}. Using an average rotation measure of $-10$ rad m$^{-2}$ \citep{https://arxiv.org/abs/2309.01716} towards the source direction, we find that the rotation measure-corrected magnetic field vectors (Fig. \ref{G166_pol_20}) follow the orientation of the relic structure seen in the high-resolution images. This alignment is characteristic of radio relics and indicates substantial magnetic field coherence in the emission region. The detection of high linear fractional polarization extending to frequencies as low as 550 MHz indicates the presence of ordered magnetic fields in the cluster outskirts, as Faraday depolarization effects become increasingly severe at these frequencies. Observational evidence \citep{2010A&A...513A..30B} and simulations \citep{2002A&A...387..383D} both show that magnetic field strength decreases with distance from the cluster center. In the outer regions, where the magnetic fields are weaker, passing shocks can more effectively modify and align these fields, resulting in the observed high linear fractional polarization. 

The linear fractional polarization spectrum (Fig. \ref{G166_mulpol}) reveals complex behavior, including abrupt variations across wavelengths that deviate from the single component depolarization models derived from higher-frequency measurements (grey lines in fig. \ref{G166_mulpol}). Notably, there is an unexpected increase in linear polarization fraction at the wavelength of 42.475 cm or the frequency of 706.3 MHz. Additionally, the depolarization curve gets flatter at the lower frequencies. To understand these features, we investigated various depolarization mechanisms, particularly external Faraday depolarization (EFD) and internal Faraday depolarization (IFD). For radio relics with typical rotation measures of $10-100~\text{rad.m}^{-2}$ (e.g. \citealp{2021ApJ...911....3D,2022A&A...657A...2R,2024A&A...691A..23D,2025ApJ...979....4P}), the structure of the RM distribution affects the observed polarization through beam depolarization effects from both internal and external Faraday dispersion \citep{2021ApJ...911....3D,2024A&A...691A..23D}).
For a source with intrinsic polarization $\text{p}_0$ and rotation measure dispersion $\sigma_{\text{RM}}$, the observed polarization follows \citep{1966MNRAS.133...67B}:
\begin{equation}
p_{\text{EFD}} = p_0 e^{-S}
\end{equation}
\begin{equation}
p_{\text{IFD}} = p_0 \frac{1 - e^{-S}}{S}
\end{equation}
where $\text{S}=2\sigma_{\text{RM}}^2\lambda^4$. Our analysis reveals that a single polarization and depolarization component cannot adequately model the observed spectrum. When the single-component model is fitted to the 1.38 and 1.71 GHz linear fractional polarization measurements, the model underpredicts (0-4\%; Fig. \ref{G166_mulpol}; grey lines) the expected linear polarization at lower frequencies from 550-750 MHz. 

We then explored two-component models incorporating two polarization and two depolarization components (e.g., \citet{https://doi.org/10.1093/pasj/psv082}), testing different combinations of internal and external Faraday depolarization components (IFD+IFD, EFD+EFD, and IFD+EFD). While these models successfully explain most of the spectrum, they fail to account for the abrupt increase in linear fractional polarization observed around 706.3 MHz. After excluding this anomalous point, we investigated various two-component model combinations.
\begin{table*}
\centering
\begin{tabular}{lcccc}
Model & \boldmath{$p_{0,1}$} & \boldmath{$\sigma_{\text{RM1}}$} & \boldmath{$p_{0,2}$} & \boldmath{$\sigma_{\text{RM2}}$} \\ \hline\hline
EFD-IFD       & $0.369 \pm 0.220$      & $36.15 \pm 10.15$                  & $0.282 \pm 0.080$    & $1.84 \pm 0.75$                 \\
IFD-IFD       & $0.349 \pm 0.221$      & $36.24 \pm 10.11$                  & $0.296 \pm 0.087$    & $2.93 \pm 1.20$                 \\
EFD-EFD       & $0.455 \pm 0.271$      & $31.53 \pm 9.70$                  & $0.373 \pm 0.079$    & $2.45 \pm 0.55$                 \\
\hline
\end{tabular}
\caption{Fitted parameters and their $1\sigma$ uncertainties for each model defined in Eq. \ref{abc}. The parameters with subscript 1 correspond to the far-side components, while those with subscript 2 correspond to the near-side components.}
\label{table}
\end{table*}
To establish the theoretical framework for our two-component analysis, we consider a geometric scenario where emission originates from two distinct regions along the line of sight. We assume two emission components with intrinsic polarization degrees $p_{0,1}$ and $p_{0,2}$, and rotation measure dispersions $\sigma_{\text{RM1}}$ and $\sigma_{\text{RM2}}$ across their respective emission regions. In this geometric configuration, the farside component to the observer undergoes depolarization once within its own emission region, then goes for another round of depolarization as it passes through the nearside component before reaching the observer. The nearside component to the observer experiences only single depolarization within its own emission region.

The measured polarization at any wavelength $\lambda$ by the observer for each of these cases are respectively:
\begin{equation}
\begin{aligned}
    p_{\text{IFD-IFD}} &= \text{IFD}(\lambda, p_{0,1}, \sigma_{\text{RM1}}) \\
    &\quad + \text{IFD}(\lambda, \text{IFD}(\lambda, p_{0,1}, \sigma_{\text{RM1}}), \sigma_{\text{RM2}}) \\
    &\quad + \text{IFD}(\lambda, p_{0,2}, \sigma_{\text{RM2}}), \\
    p_{\text{EFD-IFD}} &= \text{IFD}(\lambda, p_{0,1}, \sigma_{\text{RM1}}) \\
    &\quad + \text{EFD}(\lambda, \text{IFD}(\lambda, p_{0,1}, \sigma_{\text{RM1}}), \sigma_{\text{RM2}}) \\
    &\quad + \text{EFD}(\lambda, p_{0,2}, \sigma_{\text{RM2}}), \\
    p_{\text{EFD-EFD}} &= \text{EFD}(\lambda, p_{0,1}, \sigma_{\text{RM1}}) \\
    &\quad + \text{EFD}(\lambda, \text{EFD}(\lambda, p_{0,1}, \sigma_{\text{RM1}}), \sigma_{\text{RM2}}) \\
    &\quad + \text{EFD}(\lambda, p_{0,2}, \sigma_{\text{RM2}}).
\end{aligned}
\label{abc}
\end{equation}
We have used the a Markov Chain Monte Carlo approach provided with PyMC \citep{2015arXiv150708050S} package to fit the models to the linear fractional polarization measurements.

The stepwise decrease in polarization fraction is well-captured by the two-component models; however, the sudden increase in linear polarization remains unexplained. Although more complex multi-component models might better describe these features, the limited size of our dataset precludes such an analysis. We also considered the possibility that incomplete flux recovery due to missing UV coverage might explain the polarization jumps. However, the consistent wavelength-dependent flux increase, characteristic of radio relics, suggests this is unlikely. We acknowledge that the uniform RM assumption in Eq. \ref{abc} may be oversimplified given the spatial RM variations observed at small scales (15-30 kpcs) in other relics \citep{2021ApJ...911....3D,2022A&A...657A...2R,2024A&A...691A..23D,2025ApJ...979....4P}.

Similar step-like depolarization and abrupt jumps in linear fractional polarization spectra have been observed in radio relics within Abell 1240 \citep{2018MNRAS.478.2218H} and Abell 2256 \citep{https://doi.org/10.1093/pasj/psv082}. A detailed study of polarization properties in Abell 2256 using VLA S (2-4 GHz) and X (8-10 GHz) bands revealed comparable depolarization patterns between 1.37 and 3 GHz \citep{https://doi.org/10.1093/pasj/psv082}. Notably, they observed sudden polarization increases around 1.5 GHz, which were determined to be intrinsic features rather than artefacts of uv coverage. The southern radio relic in Abell 1240 exhibits similar behavior, with a higher polarization fraction at 1.4 GHz \citep{2018MNRAS.478.2218H} compared to 3 GHz \citep{2009A&A...494..429B}.

Single-component depolarization models have been found inadequate for some radio relics \citep{https://doi.org/10.1093/pasj/psv082,2018MNRAS.478.2218H,2019MNRAS.489.3905S,2022A&A...666A...8S}, suggesting additional complexity from three-dimensional merger shock structures projected onto the sky plane. Unlike the assumption of emission from a single Faraday depth with a Gaussian magnetic field distribution, radio relics may involve emission from multiple distinct RM components both along the line of sight and across the spatially extended regions being analyzed, each with different polarization properties \citep{2021MNRAS.500..795D,2022A&A...666A...8S,2025ApJ...979....4P}. Simulations \citep{2019MNRAS.490.3987W, 2021MNRAS.500..795D} and a few observational evidences \citep{2019MNRAS.489.3905S,2022A&A...666A...8S} indicate that magnetic field structures in relics can exhibit non-Gaussian characteristics. Depolarization across the shock region is also highly non-uniform. Emission from regions farther from the observer undergoes stronger depolarization due to additional layers of intervening magnetized plasma while front-facing shock regions are less affected. The resulting polarized emission represents a cumulative signal, integrating contributions from layers experiencing varying degrees of depolarization. Spectral index variations across the emission region introduce additional challenges, creating frequency-dependent changes in the linear fractional polarization. Furthermore, at lower frequencies, synchrotron and inverse-Compton losses lead to a more mono-energetic electron population, which can flatten polarization trends. These frequency-dependent effects suggest that the observed polarization may arise from different electron populations across the spectrum. 

Our fitted two-component depolarization models suggest a physical scenario with two distinct emission regions contributing to the observed polarization trends in radio relics. This can be thought of as, the first component originating from the far side of the shock, where polarized radiation traverses the entire shock width, undergoing more depolarization as it propagates through the turbulent, magnetized plasma of the shock region. The second component arises from the near-side shock surface as seen by the observer. While this emission is subject to depolarization from the intervening intracluster medium (ICM) which is not emitting, it avoids the depolarization effects associated with the shock width.
This two-component interpretation, accounting for different depolarization mechanisms and magnitudes, provides a more realistic representation of radio relic emission compared to single-component models. The fitted rotation measure dispersion ($\sigma_{\text{RM}}$) values (Table \ref{table}) indicate that the inner medium is more turbulent than the outer one, supporting our proposed scenario and aligning with expectations. However, the spatial averaging of polarized emission introduces additional complexities in modelling depolarization trends, and current data limitations further restrict our ability to explore multi-component depolarization scenarios.

\section{Conclusions}
In this work we have utilized the full polarization uGMRT observations of Abell 746 to study the low-frequency polarization properties of the NW radio relic. We summarize the conclusions as follows.
\begin{enumerate}[label=(\roman*)]
    \item At 650 MHz, over the polarized region, we have detected an average linear fractional polarization of $18\pm 4\%$ from the NW radio relic. When extrapolated using any single component depolarization model from the higher frequency measurements, the linear fractional polarization is underestimated.
    \item The depolarization spectrum of the NW radio relic is well-described by a two-component depolarization model excluding the exceptionally increased linear fractional polarization at the wavelength of 42.475 cm or the frequency of 706.3 MHz.
    \item The fitted parameters do not favour any specific model but consistently predict that the first depolarizing component is more turbulent compared to the second one.
    \item The radio relic depolarization spectrum is affected by spatial averaging over complex features of the radio relic emission. Currently, the number of data points across the spectrum limit exploring complex models with more than two depolarization components.
\end{enumerate}

As steep spectrum sources, radio relics are much brighter at lower frequencies. However, based on expectations from higher frequency measurements, radio relics were thought to be completely depolarized at frequencies below 1 GHz. Our work suggests that radio relics can exhibit a step-wise decrement in linear fractional polarization with decreasing frequency, and can show significant linear fractional polarization in lower frequencies, contradicting the expectations derived from higher-frequency measurements, indicating significant magnetoionic complexity within the radio relics.

\section{Acknowledgements}
We express our deep gratitude to Reinout J. van Weeren for kindly sharing the 1.38 and 1.71 GHz WSRT measurements of the Abell 746. A.P. sincerely thanks Daniel R. Wik for providing helpful comments that significantly improved the quality of the article. A.P. also acknowledges Preshanth Jagannathan and Ben Hugo for their insightful discussions about the 3C286 models at frequencies below 1 GHz. RK acknowledges the support from the SERB Women Excellence Award WEA/2021/000008. We thank the staff of the GMRT that made these observations possible. GMRT is run by the National Centre for Radio Astrophysics of the Tata Institute of Fundamental Research. This research has made use of NASA's Astrophysics Data System Bibliographic Services. This research has made use of the NASA/IPAC Extragalactic Database (NED),
which is operated by the Jet Propulsion Laboratory, California Institute of Technology,
under contract with the National Aeronautics and Space Administration.
\bibliography{sn-bibliography}
\appendix

\section{Linear fractional polarization maps}
The fractional polarization maps show the NW radio relic's polarization properties across the three observed frequencies. The 650 MHz map demonstrates polarization detection at this low frequency, while the higher frequency WSRT maps show more extended polarized emission due to reduced Faraday depolarization effects.
\begin{figure*}[ht!]
\includegraphics[width=\textwidth]{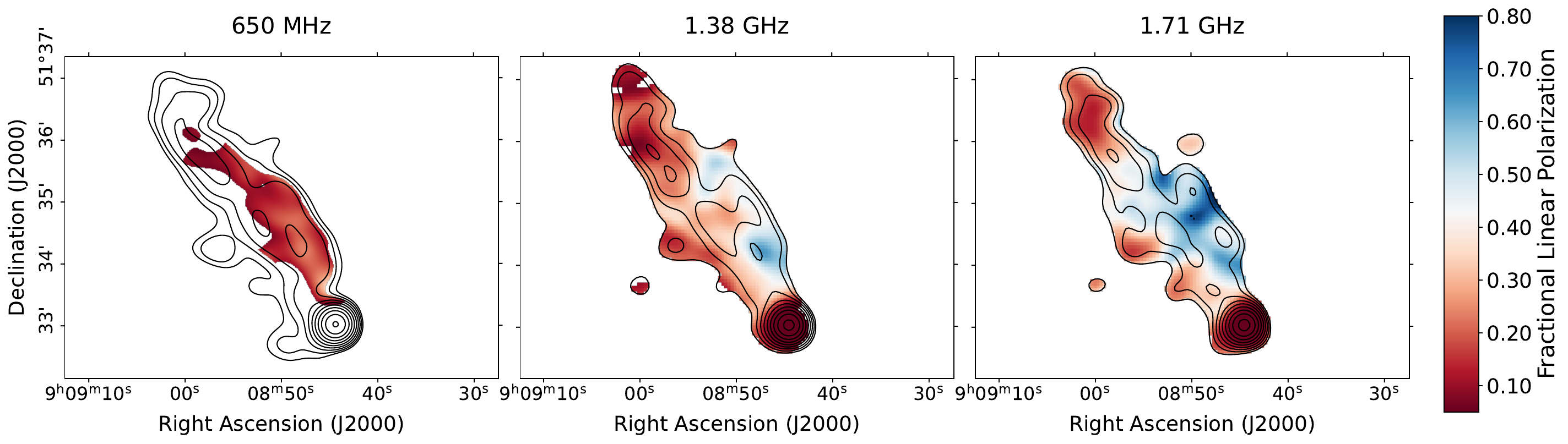}
\caption{Linear fractional polarization maps are presented at three frequencies: 650~MHz (uGMRT, left panel), 1.38~GHz (WSRT, middle panel), and 1.71~GHz (WSRT, right panel). In each case, the contours start at the 5$\sigma$ total intensity level and increase by successive factors of $\sqrt{2}$. The first contour levels are 800~$\mu$Jy~beam$^{-1}$, 300~$\mu$Jy~beam$^{-1}$, and 235~$\mu$Jy~beam$^{-1}$ at 650~MHz, 1.38~GHz, and 1.71~GHz, respectively. All images are shown at a common resolution of $25^{\prime\prime} \times 25^{\prime\prime}$.}
\label{compare}
\end{figure*} 
\end{document}